# Exciton-related electroluminescence from monolayer $MoS_2$

Yu Ye[1,†], Ziliang Ye[1,†], Majid Gharghi[1], Hanyu Zhu[1], Mervin Zhao[1], Xiaobo Yin[1,2] & Xiang Zhang[1,2,*]

[1] NSF Nanoscale Science and Engineering Center, 3112 Etcheverry Hall, University of California, Berkeley, California 94720, USA.

[2] Materials Sciences Division, Lawrence Berkeley National Laboratory, 1 Cyclotron Road, Berkeley, California 94720, USA.

[†] These authors contributed equally to this work.

[*] Correspondence and requests for materials should be addressed to X. Z. (email: xiang@berkeley.edu)

**Abstract**

Excitons in $MoS_2$ dominate the absorption and emission properties of the two-dimensional system. Here, we study the microscopic origin of the electroluminescence from monolayer $MoS_2$ fabricated on a heavily *p*-type doped silicon substrate. By comparing the photoluminescence and electroluminescence of a $MoS_2$ diode, direct-exciton and bound-exciton related recombination processes can be identified. Auger recombination of the exciton-exciton annihilation of bound exciton emission is observed under a high electron-hole pair injection rate at room temperature. We expect the direct exciton-exciton annihilation lifetime to exceed the carrier lifetime, due to the absence of any noticeable direct exciton saturation. We believe that our method of electrical injection opens a new route to understand the microscopic nature of the exciton recombination and facilitate the control of valley and spin excitation in $MoS_2$.

## Introduction

The direct energy bandgap and the non-centrosymmetric lattice structure set monolayer MoS$_2$ distinct from its bulk counterpart and the widely studied monolayer graphene[1-3]. Strong photoluminescence, microscopic mechanisms of exciton-related recombination, large exciton binding energy and the possibility for efficient control of valley and spin have attracted extensive research effort into this material[4-10]. Prior research has demonstrated that excitons dominate the emission properties of these two-dimensional systems[11, 12]. Electroluminescence, i.e. photon emission from radiative recombination of the electrically injected electrons and holes, is a reliable way to study exciton recombination processes in monolayer MoS$_2$, including valley and spin excitation and control. Past work suggests that monolayer MoS$_2$ has a potential as a two-dimensional light emitter[13], in which the electroluminescence occurs through a hot carrier process and is localized in the region adjacent to contacts. However, the low electroluminescent efficiency and signal-to-noise ratio obscure the understanding of contributions of individual optical transitions. Here, we report electrically pumped light emission from heterojunction of monolayer MoS$_2$ (*n*-type) and heavily doped (*p*-type) silicon. A new level of control over electrical carrier injection is achieved, resulting in high signal-to-noise ratio emission spectrum, allowing for the identification of emission from different optical transitions.

In our heterojuncion diodes (Fig. 1a), heavily *p*-doped silicon is used to inject holes to *n*-type monolayer MoS$_2$. A pattern of highly doped *p*-type silicon on silicon oxide was microfabricated to create a step sidewall. We developed a site-control transfer method to place the monolayer MoS$_2$ across the silicon/silicon oxide step. Figure 1b shows the ideal corresponding band structure under forward bias; the built-in potential and applied voltage are supported by a depletion layer with abrupt boundaries, and outside the boundaries the semiconductor is assumed to be neutral. When a forward bias is applied to the heterojunction, the injection of holes from silicon across the junction can give rise to efficient radiative recombination, because of direct band-gap property of monolayer $MoS_2$. The corresponding electroluminescence will be determined by the radiative transition of the monolayer $MoS_2$. Practically, due to the large valence band offset between silicon and monolayer $MoS_2$[14], the energy band of $MoS_2$ will bend upward under high forward bias, which is not favorable for hole-injection from silicon. In Fig. 1c we present the *I-V* characteristic of a monolayer $MoS_2$ diode, which clearly shows rectifying behavior when the voltage changes from 3 V to –3 V.

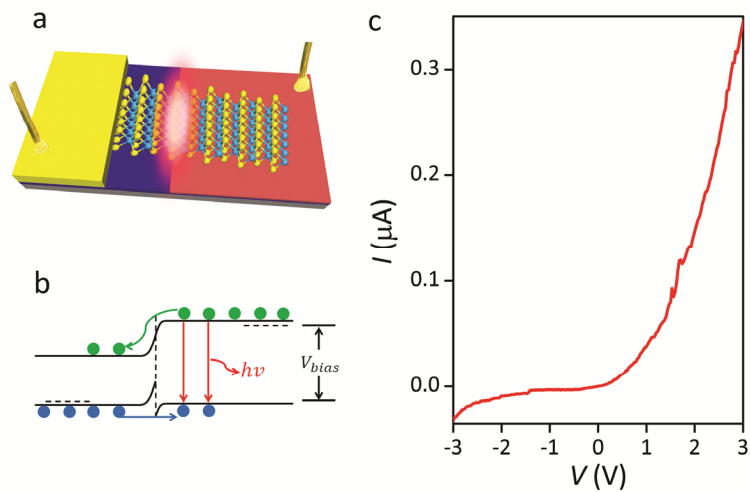

**Figure 1. Device structure and electronic characteristics.** a, Schematic of the $MoS_2$/silicon heterojunction electroluminescence device. b, Ideal band structure of the $MoS_2$ diode under forward bias. c, Electrical characteristics of the $MoS_2$ diode.

Figure 2a depicts the electroluminescent emission captured by a single-photon sensitive camera (Andor DL-604M-#VP) from a device at forward bias voltage of 5 V and current of 42 µA at room temperature. After superimposing a white light scattering image of the device, we find that the electroluminescence is localized at the edge of the heterojunction. By applying an in-plane bias voltage, the largest voltage drop naturally occurs across the heterojunction edge due to the semiconducting characteristics of $MoS_2$[15]. This is further confirmed by the electrostatic potential mapping by a scanning photocurrent microscopy (not shown here).

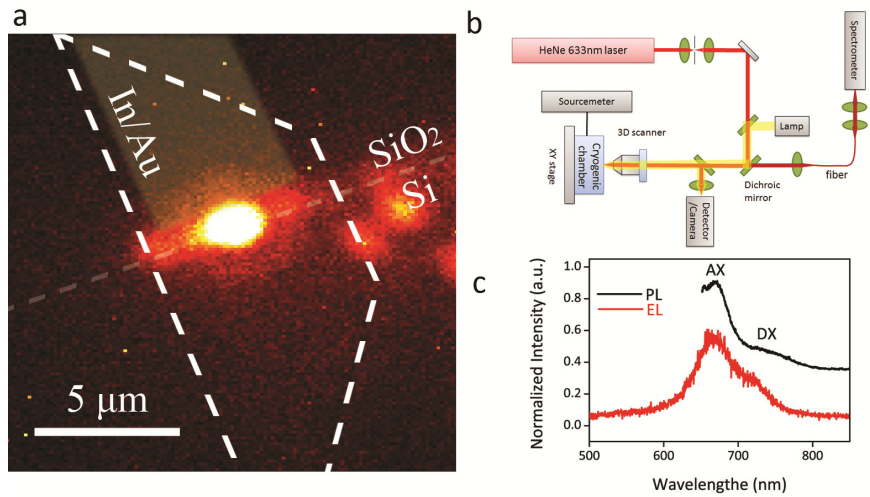

**Figure 2. Identification of the light emission and the setup.** a, Surface plot of the electroluminescent emission. A white light scattering image of the device is overlapped to verify that the emission is localized at the edge of the heterojunction. The white dash line indicates the region of the monolayer $MoS_2$. Scale bar: 5 µm. b, Schematic of our optical setup. The emission is collected through a 50x objective with 0.55 NA. c, Photoluminescence and eletroluminescnece spectra from the same device. The electrolunminence spectrum is measured at the current of 42 µA. Besides the strong direct exciton emission (labelled AX), a weaker satellite peak at a lower energy is observed (labelled DX). They are attributed to A exciton and bound exciton emission.

To measure the photoluminescence and electroluminescence spectra, we coupled the emitted light to a fiber connected to a spectrometer (Andor Shamrock 303) as shown in Fig. 2b. Figure 2c shows the room-temperature electroluminescence spectrum of monolayer $MoS_2$ at a current of 42 µA. The two principal luminescence features at 667 nm (labeled AX) and 720 nm (labeled DX) are associated with the A exciton and the bound exciton of monolayer $MoS_2$, respectively. Compared with the photoluminescence spectrum plotted in the same figure (Fig. 2c), the high electrical bias causes spectral broadening; thermal background also increases in the electroluminescence spectrum. The B exciton, which has a higher energy close to that of the A exciton[6], is indistinguishable in the electroluminescence due to low efficiency of the B exciton process and the broadening of the electroluminescence[13]. A and B are direct excitons in $MoS_2$ with the energy split from valence band spin-orbital coupling.

Figure 3a depicts the room-temperature electroluminescence spectra under varied current. The electroluminescence exhibits a current threshold of about 15 µA in this device; the two main features of the AX and DX excitation emissions are clearly observed at currents exceeding this threshold. They can be well fitted with a two-Lorentzian model. In Fig. 3c, we present the current dependence of the AX and DX emission intensities as extracted from Fig. 3a. The A exciton emission, AX, shows a linear dependence with increasing current. However, the bound exciton, DX, rises linearly at low currents but saturates as the current exceeds about 65 µA. Saturation of the DX exciton emission cannot be caused by a phonon-assisted nonradiative process, as the two peaks display different current dependencies. We propose this is a different effect involving multiple exciton-exciton interactions, which is similar to Auger recombination, a process well documented in tightly confined carbon nanotube systems. Auger recombination may lead to rapid exciton-exciton annihilation when extra excitons or multiple excitations are present[16-21]. At a low electron-hole pair injection rate of $I/2q$ ($q$ as the electron charge), there may only be one electron-hole pair in excited monolayer $MoS_2$. Thus, we observe linear dependence with increasing current at low injection rate. When the electron-hole pair injection rate exceeds the inverse carrier lifetime $\tau_L^{-1}$, more than one electron-hole pair is present in monolayer $MoS_2$. The Auger process opens up a nonradiative recombination channel for electron-hole pair recombination. If the Auger process is sufficiently efficient, it will quickly deplete the population of electron-hole pairs. The annihilation of the electron-hole pairs comes to a stop when only a single electron-hole pair remains in the monolayer $MoS_2$. Thus, we observe the saturation of the exciton emission at sufficiently high injection current. The sudden saturation further suggest that the DX-DX annihilation lifetime $\tau_A^{DX} \ll \tau_L$. On the other hand, due to the absence of any noticeable AX saturation, we expect the AX-AX annihilation lifetime $\tau_A^{AX}$ to be much longer than $\tau_L$.

It is well known the microscopic mechanism of optical transition is temperature dependent[12]. When the device temperature is reduced to 10 K, the electroluminescence resonance exhibited a blue-shift from 667 nm to 662 nm for the AX peak and from 720 nm to 701 nm for DX peak, while the full-width at half-maximum linewidth decreased down to 32 nm. The electroluminescence spectrum is very consistent with the photoluminescence spectrum at low temperature. Deviations are caused by a slight red-shift and spectral broadening arising from the high current induced local temperature rise.

Again, the two main features of the AX and DX excitation emissions can be clearly read out under all the currents above threshold (Fig. 3b). In Fig. 3d, we present the current dependence of the AX and DX emission intensities as extracted from Fig. 3b. Surprisingly, the DX emission exhibits different current dependence behavior compared to that of room temperature. Both AX and DX emissions show a linear increase with current. The absence of the saturation of the DX emission under a high electron-hole injection rate at low temperature is due to the slowing down of the Auger rate[21]. Past work has already demonstrated that the lifetime of the electron-hole pair recombination at low temperature (4.5 K) is prolonged by more than one order compared to that of room temperature[19, 21]. As the increased DX-DX annihilation lifetime $\tau_A^{DX}$ is comparable or longer than $\tau_L$, the DX emission will not saturate even under high injection rate. To the best of our knowledge, our experiment demonstrates the first observation of Auger recombination of the exciton-exciton annihilation of the DX emission in monolayer $MoS_2$ system. Further experimental and theoretical work will focus on measuring the timescale of the Auger recombination process.

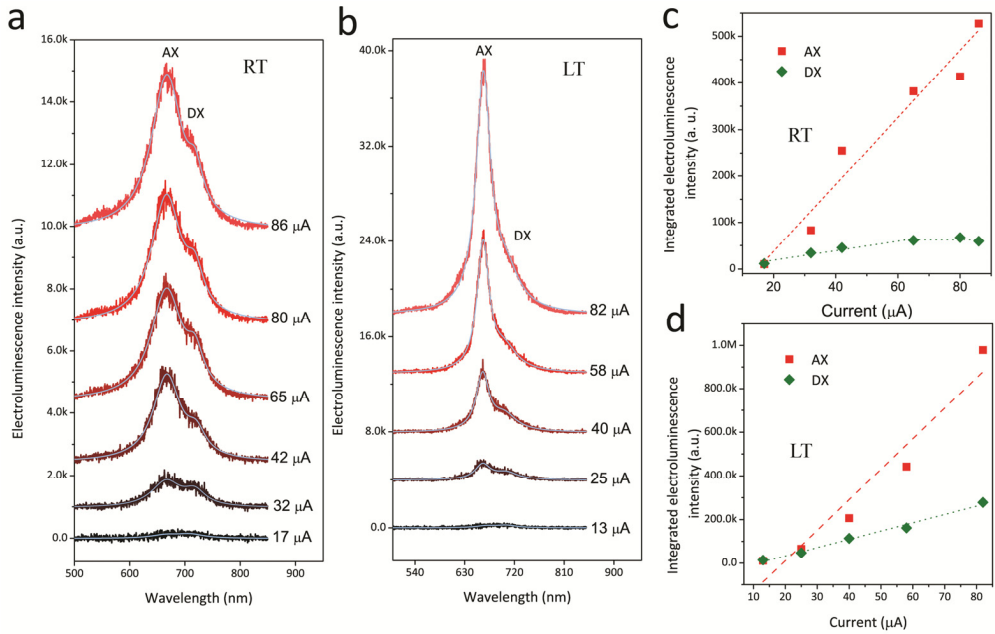

**Figure 3. Electroluminescence spectra.** a, Room-temperature electroluminescence spectra of a $MoS_2$ diode recorded at different currents levels, ranging from 17 to 86

µA. The data can be fitted well with two Lorentz contributions, which are attributed to A exciton (labelled AX) and bound exciton (labelled DX) emission. b, 10 K electroluminescence spectra of the MoS$_2$ diode recorded at different current levels, ranging from 13 to 82 µA. The data can be fitted well with two Lorentz contributions, which are attributed to A exciton (labelled AX) and bound exciton (labelled DX) emission. c, Red symbols: the A exciton emission (AX) shows an approximately linear increase with current. Green symbols: bound exciton emission (DX). The electroluminescence saturates as the current exceed ~65 µA at room temperature. The dashed lines are fitting results. d, Red symbols: the A exciton emission (AX) shows an approximately linear increase with current. Green symbols: bound exciton emission (DX) also shows an approximately linear increase with current at low temperature. The dashed lines are fitting results.

**Conclusion**

In conclusion, we report the electroluminescence of monolayer MoS$_2$ fabricated on a heavily *p*-type doped silicon substrate. The high signal-to-noise ratio allows for the identification of emission from different optical transitions. Auger recombination of the exciton-exciton annihilation of bound exciton emission is observed under a high electron-hole pair injection rate at room temperature for the first time. Further experimental and theoretical work is needed to identify the microscopic nature of the exciton recombination.

**Acknowledgements**

Y. Ye would like to thank T. Cao for helpful discussion.

**Author contributions**

Y. Y. and Z. Y. were responsible for the experimental work. M. G. and H. Z.

assisted Y. Y. fabricate the devices. X. Y. and X. Z instructed the project. All authors discussed the results and commented on the manuscript.